\documentclass[12pt]{article}

\makeatletter

\newcommand{\LyX}{L\kern-.1667em\lower.25em\hbox{Y}\kern-.125emX\@}

\newcommand{\lyxaddress}[1]{
  \par {\raggedright #1 
  \vspace{1.4em}
  \noindent\par}
}

\makeatother

\begin{document}

\thispagestyle{plain}
\title{Correlator mixing and mass reduction \\
as signals of chiral symmetry restoration}

\maketitle
\medskip{}

\begin{center}

\author{J. Delorme\protect\protect\( ^{1}\protect \protect \), M. Ericson\protect\protect\( ^{1,2,4}\protect \protect \)
, P.A.M. Guichon\protect\protect\( ^{3}\protect \protect \), A.W. Thomas\protect\protect\( ^{4}\protect \protect \)}

\vskip 15pt

\lyxaddress{{\footnotesize \protect\protect\( ^{1}\protect \protect \)}\emph{\footnotesize IPNLyon,
IN2P3-CNRS et UCB Lyon I, 43 Bvd du 11 Novembre 1918, F69622 Villeurbanne Cedex}\footnotesize }

\vspace{-10pt}

\lyxaddress{{\footnotesize \protect\protect\( ^{2}\protect \protect \)}\emph{\footnotesize Theory
division, CERN, CH12111 Geneva}\footnotesize }

\vspace{-10pt}

\lyxaddress{{\footnotesize \protect\protect\( ^{3}\protect \protect \)}\emph{\footnotesize SPhN/DAPNIA,
CEA-Saclay, F91191 Gif sur Yvette Cedex}\footnotesize }

\vspace{-10pt}

\lyxaddress{{\footnotesize \protect\protect\( ^{4}\protect \protect \)} \emph{\footnotesize Department
of Physics and Mathematical Physics and Special Research Center for the Subatomic
Structure of Matter, University of Adelaide, SA 5005, Australia}\footnotesize }

\end{center}

\begin{abstract}
Chiral symmetry restoration in a dense medium is to some extent a consequence
of the nuclear pion cloud. These pions induce a mixing of the axial and vector
current contributions in the axial and vector correlators. We discuss their
influence on hadron masses and investigate the signal produced by the remaining
contribution associated with chiral symmetry restoration. Using the quark-meson
coupling model we find that the latter is associated with the reduction of hadron
masses. \bigskip{}

\end{abstract}
\vfill{\textit{CERN-TH/99-107}\\
\textit{\scriptsize LYCEN 9935}\textit{\small ,} \textit{\scriptsize
DAPNIA/SPhN-99-25,
ADP-99-20/T362}\textit{\small  }}{\small \par}
\newpage

Chiral symmetry is partially restored in a heat bath or in dense matter. For
instance, at normal nuclear matter density the quark condensate, which is the
order parameter of the spontaneous breaking of the symmetry, has decreased by
about 35\%. It is therefore legitimate to investigate whether this large reduction
is accompanied by precursor effects which signal the approach to full restoration.
These effects should be linked to the symmetry itself. The realization of chiral
symmetry in the Wigner rather than in the Goldstone mode implies either a vanishing
of all hadron masses or the existence of parity doublets, each state being degenerate
with its chiral partner. The fact that neither of these situations occurs for
free hadrons was the motivation to favour the Goldstone realization. The signatures
of the Wigner realizations make it natural to look for precursor effects of
the full restoration in the form of a decrease of the hadron masses or in some
effects which involve parity. The first road was taken by Brown and Rho~\cite{BR}
on the basis of scale invariance arguments. They suggested a reduction of hadron
masses linked to the condensate evolution. For tests of these ideas attention
has focussed on the \( \rho  \) and \( \omega  \) masses. On the other hand,
Dey \textit{et al.}~\cite{DEY} have shown that in a heat bath there will be
some mixing of axial and vector correlators. This mixing is induced by the emission
or absorption of an s-wave thermal pion which produces a change in the parity.
For soft pions, \textit{i.e.} in the chiral limit and at low temperatures, the
amount of axial (vector) correlator introduced in the vector (axial) one is
governed by the condensate evolution. At the same time the original correlator
is depleted by the same amount.

A similar concept was introduced by Chanfray \textit{et al.}~\cite{CDE} in
dense matter. The pions in this case are the virtual ones of the nuclear pion
cloud. Since they constitute an integral part of the nucleus there is no true
correlator mixing. Nevertheless Chanfray \textit{et al.} extended the notion
of mixing to the currents in such a way that they could build a coherent picture
from the nuclear pions. Similar to the case of the heat bath, there occurs a
depletion phenomenon in the form of a quenching of certain coupling constants
from the nuclear pion loops. The quenching factor is related to the pion scalar
density. One difference from the case of a heat bath is that the relation between
this quantity and the condensate is not straightforward. The condensate evolution
in a dense medium is governed by the nucleon sigma commutator \( \Sigma _{N} \).
For a uniform medium with nucleon scalar density \( \rho _{s} \), the condensate
evolves, to first order in \( \rho _{s} \), according to the relation: 
\begin{equation}
\label{eq1}
\frac{<\overline{\psi }\psi (\rho _{s})>}{<\overline{\psi }\psi (\rho _{s}=0)>}=1-\frac{\Sigma _{N}\rho _{s}}{f_{\pi }^{2}m_{\pi }^{2}}\, \, .
\end{equation}
 The question is then how much of the restoration arises from the pion cloud.

It turns out that our question concerning the role of the pion cloud in the
restoration of chiral symmetry in dense matter cannot be given in a model independent
way. For example, as we explain later, within the cloudy bag model the nucleon
sigma commutator separates into a pionic contribution and a remaining piece~\cite{SIGC},
\( \Sigma ^{r}_{N} \) : 
\begin{equation}
\label{eq2}
\Sigma _{N}=\frac{m_{\pi }^{2}}{2}\int d\vec{x}<N|\phi ^{2}(\vec{x})|N>+\Sigma _{N}^{r}\, \, .
\end{equation}
 We introduce the (constant) average quantity, \( <\phi ^{2}> \), related to
the pion scalar density \( \rho ^{s}_{\pi }=m_{\pi }<\phi ^{2}>\, \,  \): 
\begin{equation}
\label{eq2'}
<\phi ^{2}>=\rho _{s}\int d\vec{x}<N|\phi ^{2}(\vec{x})|N>\, \, .
\end{equation}
 The symmetry restoration is produced by both pionic and non-pionic terms on
the r.h.s. of Eq.(\ref{eq2}) according to 
\begin{equation}
\label{eq2''}
\frac{<\overline{\psi }\psi (\rho _{s})>}{<\overline{\psi }\psi (\rho _{s}=0)>}=1-\frac{<\phi ^{2}>}{2f_{\pi }^{2}}-\frac{\rho _{s}\Sigma _{N}^{r}}{f_{\pi }^{2}m_{\pi }^{2}}\, \, .
\end{equation}
 A similar expression holds for the evolution with temperature in the heat bath.
However, in that case only the term in \( <\phi ^{2}> \) is present in the
r.h.s. of Eq.~(\ref{eq2''})~\cite{CEW}.

As a matter of fact the pions are responsible for the correlator mixing which
signals chiral symmetry restoration. The question is whether they also affect
the nucleon mass and what is the restoration signal due to \( \Sigma _{N}^{r} \).
The evolution of the hadron masses at finite temperature has been discussed
by several authors (see e.g., Refs.~\cite{LS,KZ}). Here we concentrate on the
modification which is directly linked to the evolution of the condensate, that
is to the existence of a non zero expectation value for \( <\phi ^{2}>. \)
This expectation value can arise in either nuclear matter or a heat bath. In
the latter case \( <\phi ^{2}> \) goes like \( T^{2} \) in the chiral limit.
Using the effective Lagrangian of Lynn, in the tree approximation~\cite{LYN,DCE},
it is straightforward to derive the following expression for the change in the
nucleon mass (consistent with the results of Ref.~\cite{KZ}): 
\begin{equation}
\label{eq2'''}
\frac{\Delta M^{s}}{M}=-\frac{\Sigma _{N}}{M}\frac{<\phi ^{2}>}{2f_{\pi }^{2}}\, \, .
\end{equation}
 We have to specify the meaning of the index \textit{s}, in Eq.(\ref{eq2'''}),
which stands for soft. The pions which build the quantity \( <\phi ^{2}> \)
need not be soft, and indeed the virtual nuclear pions are not, with their typical
momentum being a few hundred MeV/c. However, the term in \( <\phi ^{2}> \)
of Eq.~(\ref{eq2''}) is the only one to survive in the soft limit. Practically,
in the N-N interaction, the two-pion exchange with a contact s-wave \( \pi \pi NN \)
coupling which leads to the mass modification of Eq.(\ref{eq2'''}) represents
a minor contribution ~\cite{DCE}. On the other hand, for non-soft pions the
derivative couplings introduce other terms in the mass modification. The extra
terms will be lumped into an effective \( \sigma  \) meson exchange as explained
below. An expression similar to Eq.~(\ref{eq2'''}) holds for the other hadrons.
By comparison with the evolution of the condensate of pionic origin, Eq.~(\ref{eq2''}),
the mass evolution is thus attenuated by the factor \( \Sigma _{N}/M \) which
vanishes in the chiral limit as imposed by the soft pion theorems. Thus in this
limit there is no mass shift of order \( T^{2} \), as we know on general grounds~\cite{LS,EI},
and in the nuclear case no term of order \( m_{\pi } \) appears in the mass
change~\cite{BIR}.

Coming now to the restoration signal due to \( \Sigma _{N}^{r} \), in order
to be definite, we use the quark-meson coupling model (QMC) of Guichon~\cite{PAM,QMC}
where the quarks interact locally with \( \sigma  \) and \( \omega  \) fields.
This phenomenological \( \sigma  \) exchange, when taken between two-nucleon
states, incorporates among other things the excitation via p-wave pions of one
or both nucleons into \( \Delta  \) resonances. Therefore it also includes,
in a very phenomenological fashion, the influence of non-soft pions on the nucleon
mass beyond the modification described in Eq.~(\ref{eq2'''}). This is sufficient
for our purpose. Our aim here is not a detailed quantitative description of
the total mass change. We want instead to understand what are the signatures
of chiral symmetry restoration induced by its different components displayed
in Eq.~(\ref{eq2''}).

Originally the QMC model was devised without any reference to chiral symmetry.
The reason was that in the mean field approximation the pion field vanishes
in nuclear matter. But in order to study chiral symmetry restoration one clearly
needs a model in which the explicit breaking of chiral symmetry vanishes with
the current quark mass \( m_{q} \). For this we start from the cloudy bag model
~\cite{CBM} which is the MIT bag model where the confining surface term \( 1/2\delta (S)\overline{q}q \)
is made chirally symmetric by a coupling to a pion field \( \vec{\phi } \).
We note \( q \) the quark field of the model so as to avoid any confusion with
the QCD quark field \( \psi . \) In the following we never need to identify
\( q \) with \( \psi  \) , which would be unjustified. Our only hypothesis
is that we can use this model to evaluate \( \Sigma _{N}. \) The Lagrangian
density is: 
\begin{equation}
\label{eq3}
{\mathcal{L}}=\Theta (V)\left[ \frac{i}{2}\overline{q}\gamma ^{\mu }.(\overrightarrow{\partial }_{\mu }-\overleftarrow{\partial }_{\mu })q-m_{q}\overline{q}q-B\right] -\frac{1}{2}\delta (S)\overline{q}U^{2}(\phi )q+{\mathcal{L}}(\phi )
\end{equation}
 with 
\begin{equation}
\label{eq4}
U(\phi )=\exp (i\vec{\tau }.\vec{\phi }\gamma _{5}/2f_{\pi })\, \, 
\end{equation}
 and \( {\mathcal{L}}(\phi ) \) the Lagrangian density for the pion field 
\begin{equation}
\label{eq4'}
{\mathcal{L}}(\phi )=\frac{1}{2}D_{\mu }\vec{\phi }D^{\mu }\vec{\phi }+\frac{1}{2}m_{\pi }^{2}\phi ^{2}\, \, .
\end{equation}
 The covariant derivative on the pion field is defined as 
\begin{equation}
\label{eq4''}
D_{\mu }\vec{\phi }=\hat{\phi }\partial _{\mu }\phi +f_{\pi }\sin (\phi /f_{\pi })\partial _{\mu }\hat{\phi }\, \, .
\end{equation}
 Under infinitesimal \( SU(2)\times SU(2) \) transformations, with parameters
\( (\vec{\epsilon },\vec{\epsilon }_{5}) \), the quark field \( q \) transforms
as 
\begin{eqnarray}
q & \rightarrow  & (1+\frac{i\vec{\epsilon }.\vec{\tau }}{2})q,\label{eq5} \\
q & \rightarrow  & (1+i\gamma _{5}\frac{\vec{\epsilon }_{5}.\vec{\tau }}{2})q\, \, ,\label{eq6} 
\end{eqnarray}
 while the corresponding transformations for the pion field are 
\begin{equation}
\label{eq7}
\vec{\phi }\rightarrow \vec{\phi }-\vec{\epsilon }\times \vec{\phi },
\end{equation}
 and 
\begin{eqnarray}
\cos (\phi /f_{\pi }) & \rightarrow  & \cos (\phi /f_{\pi })+\vec{\epsilon }_{5}.\hat{\phi }\sin (\phi /f_{\pi })\, \, ,\label{eq8} \\
\hat{\phi }\sin (\phi /f_{\pi }) & \rightarrow  & \hat{\phi }\sin (\phi /f_{\pi })-\vec{\epsilon }_{5}\cos (\phi /f_{\pi }),\label{eq8'} 
\end{eqnarray}
 where Eqs.~(\ref{eq8}, \ref{eq8'}) are for the axial transformations. It
is straightforward to establish that the Lagrangian density of Eq.(\ref{eq3})
is chirally invariant under these transformations, except for the term \( m_{q}\overline{q}q\Theta (V)-m_{\pi }^{2}\phi ^{2}/2 \)
, which vanishes in the chiral limit. Using the equations of motion it is straightforward
to derive the expression for the double commutator 
\begin{equation}
\label{eq9}
[Q_{5},\frac{d}{dt}Q_{5}]=\int d\vec{x}\left[ m_{q}\overline{q}q\Theta (V)+\frac{1}{2}m_{\pi }^{2}\phi ^{2}\right] \, \, ,
\end{equation}
 which shows again that chiral symmetry is correctly implemented in the model.
By taking the expectation value of Eq.(\ref{eq9}) we identify the non-pionic
contribution to the nucleon sigma commutator: 
\begin{equation}
\label{eq9'}
\Sigma _{N}^{r}=m_{q}\int d\vec{x}<N|\Theta (V)\overline{q}q(\vec{x})|N>\, \, ,
\end{equation}
 Despite successful phenomenological application of the model defined by Eq.(\ref{eq3}),
it was soon realised that this formulation was not convenient. The reason is
that the \( \gamma _{5} \) coupling of the pion at the surface produces a strong
coupling to the negative energy quark states. Ignoring this coupling is allowed
in some cases but in order to recover the low energy theorems one has actually
to sum over all those states. This admixture of negative energy quark states
obscures the simple interpretation of the nucleon as a three valence quarks
(in first approximation) . To remedy this problem one introduces a new quark
field \( Q \) according to~\cite{AWT}
\begin{equation}
\label{eq10}
Q=U(\phi )q\, \, .
\end{equation}
 With this new field the surface coupling of the pion is shifted to a volume
coupling and the Lagrangian density becomes 
\begin{equation}
\label{eq11}
{\mathcal{L}}=\Theta (V)\left[ \frac{i}{2}\overline{Q}\gamma ^{\mu }.(\overrightarrow{\partial }_{\mu }-\overleftarrow{\partial }_{\mu })Q-m_{q}\overline{Q}U(\phi )^{\dagger 2}Q-B\right] -\frac{1}{2}\delta (S)\overline{Q}Q+{\mathcal{L}_{\pi N}}+{\mathcal{L}}(\phi ),
\end{equation}
 where the interaction term 
\begin{equation}
\label{eq12}
{\mathcal{L}_{\pi N}}=\Theta (V)i\overline{Q}\gamma ^{\mu }\left[ U(\phi )\partial _{\mu }U^{\dagger }(\phi )\right] Q
\end{equation}
 generates all the soft pion theorems if one takes its expectation value between
pion-nucleon states~\cite{AWT}. One then can check that the coupling to the
negative energy states is suppressed by powers of the pion field energy, which
supports the interpretation that the nucleon is approximately described by three
valence quarks in the lowest energy mode of the field \( Q \). In other words,
as long as we consider nucleon matrix elements, the field \( Q \) is approximately
independent of \( \phi  \). More precisely \( Q \) is not affected by the
soft content of \( \phi  \).

We are now in a position to introduce the coupling to the \( \sigma  \) and
\( \omega  \) in a way consistent with chiral symmetry. Since the combinations
\( \overline{Q}Q \) and \( \overline{Q}\gamma _{\mu }Q \) are obviously invariant
under a chiral transformation, we can add to \( {\mathcal{L}} \) the term 
\begin{equation}
\label{eq13}
{\mathcal{L}_{\sigma \omega }}=\Theta (V)\left[ -g_{\sigma }^{Q}\sigma \overline{Q}Q+g_{\omega }^{Q}\omega _{\mu }\overline{Q}\gamma ^{\mu }Q\right] +{\mathcal{L}^{0}_{\sigma \omega }}
\end{equation}
 where \( {\mathcal{L}^{0}_{\sigma \omega }} \) is the usual Lagrangian density
for free scalar and vector fields. The \( \omega  \) field, which plays no
role in the following discussion, is mentioned here only for completeness. We
stress that the \( \sigma  \) field just introduced has nothing to do with
the chiral partner of the pion field in the linear sigma model, it is a chiral
singlet and the interaction in Eq.(\ref{eq13}) is chirally invariant. From
Eq.~(\ref{eq13}) we get the following equation for the sigma field: 
\begin{equation}
\label{eq14}
(\frac{\partial ^{2}}{\partial t^{2}}-\nabla ^{2}+m_{\sigma }^{2})\sigma (x)=g_{\sigma }^{Q}\overline{Q}Q(x)\: \Theta (V)
\end{equation}
 where the r.h.s of Eq.(\ref{eq14}) is the scalar source for a single bag located
inside the volume \( V \). In a uniform nuclear medium, and in the absence
of a response of the bag, the \( \sigma  \) field has the constant value 
\begin{equation}
\label{eq15}
\overline{\sigma }=\frac{g_{\sigma }^{Q}\rho _{s}}{m_{\sigma }^{2}}\int d\vec{x}<N|\Theta (V)\overline{Q}Q(\vec{x})|N>=\frac{g_{\sigma }^{N}\rho _{s}}{m_{\sigma }^{2}}\, \, ,
\end{equation}
 where the second equation defines the \( \sigma  \)-nucleon coupling constant
\( g_{\sigma }^{N} \). To leading order, which is enough for our considerations,
the sigma field governs the change in the nucleon mass according to 
\begin{equation}
\label{eq16}
\Delta M=-g_{\sigma }^{N}\overline{\sigma }\, \, .
\end{equation}
 Notice that, contrary to the \( \sigma  \) field of the non-linear sigma model
which is constrained by the chiral circle, \( \overline{\sigma } \) (which
as already mentioned is not related to the chiral partner of the pion), can
develop large values. For instance, in the QMC model the favoured values of
the coupling constants lead to a mass change for the nucleon, \( \Delta M\sim -200{\textrm{MeV}}\, \,  \),
at normal nuclear matter density~\cite{QMC}. The corresponding number for the
vector mesons is of order -140 MeV~\cite{GUI}.

We now wish to relate the change of the nucleon mass to the condensate evolution
governed by \( \Sigma _{N}^{r} \) as defined by Eqs.~(\ref{eq2''}, \ref{eq9'}).
From Eqs.~(\ref{eq4}, \ref{eq10}), and for a homogeneous medium, in which
\( <\phi ^{2}>\, \,  \) is constant, we have to lowest order in \( <\phi ^{2}>/f_{\pi }^{2}\, \,  \):
\begin{equation}
\label{eq17}
\int d\vec{x}<N|\Theta (V)\overline{q}q(\vec{x})|N>=(1-\frac{<\phi ^{2}>}{2f_{\pi }^{2}})\int d\vec{x}<N|\Theta (V)\overline{Q}Q(\vec{x})|N>\, \, .
\end{equation}
 Here we emphasise that it is \( Q \) which is independent of \( \phi . \)
The modification of the nucleon mass is thus 
\begin{equation}
\label{eq18}
\Delta M=-\frac{g_{\sigma }^{N}g_{\sigma }^{Q}}{m_{\sigma }^{2}}\frac{\rho _{s}\int d\vec{x}<N|\overline{q}q(\vec{x})|N>}{1-<\phi ^{2}>/(2f_{\pi }^{2})}\, \, .
\end{equation}
 Using Eqs.~(\ref{eq2''},\ref{eq9'}) and the Gell-Mann-Oakes-Renner relation,
this can be written in terms of the condensate evolution, leading to 
\begin{equation}
\label{eq19}
\Delta M=-\frac{g_{\sigma }^{N}g_{\sigma }^{Q}}{m_{\sigma }^{2}}\left[ \frac{<\overline{\psi }\psi (\rho _{s})>}{1-<\phi ^{2}>/(2f_{\pi }^{2})}-<\overline{\psi }\psi (\rho _{s}=0)>\right] \, \, ,
\end{equation}
 valid up to lowest order in \( <\phi ^{2}> \). In comparison with the condensate
evolution we have the additional term in denominator involving \( <\phi ^{2}>\, \, . \)
Its presence guarantees that the modification of the mass of the nucleon is
independent of the quantity \( <\phi ^{2}>\, \, . \) It is quite remarkable
that the mass evolution, in spite of the chiral invariant coupling of the sigma
field, follows the condensate evolution, but only that piece of it which is
of non-pionic origin. We have derived this result within the QMC model where
the sigma field is coupled in a chiral invariant way to the quarks. It is likely
to be more general. The general arguments based on scale invariance~\cite{BR}
do not seem to distinguish between the different origins of the symmetry restoration
as we do.

Note that in the QMC model the proportionality constant between the mass and
the condensate is purely phenomenological. On the other hand, within the Nambu-Jona-Lasinio
(NJL) model in its simplest version, the nucleon mass is taken as three times
the effective quark mass. In the standard treatment of the model, which amounts
to the Hartree approximation, the quark mass is obtained from a gap equation
which does not incorporate meson loops and thus follows the condensate in the
chiral limit. The question is how to extend the model in such a way that the
pion density influences the mass in a negligible way while keeping a large role
in the condensate -- which amounts to preserving the soft pion theorems. The
understanding of this point, following, for example, the extension of the NJL
model to meson loops developed in Ref.~\cite{RIP}, would provide in a QCD inspired
theory, the link between the mass and the condensate which respects the constraints
of chiral perturbation theory.

In conclusion, with the results obtained here we have achieved a satisfactory
description of the signatures of the partial restoration of chiral symmetry
in a dense medium. Part of the restoration arises from the finite value of the
average squared pion field (equivalently the scalar pion density). The signature
for that part is the depletion of the axial and vector correlators and the generalized
mixing of these correlators in the sense defined in Ref.~\cite{CDE}. The mass
reduction associated with the pion density is very small, being damped by factors
of order \( m_{\pi }^{2} \) as required by low energy theorems. On the other
hand, another part of the restoration is not linked to the pion density and
does not induce any mixing. We have shown, within the QMC model, that it is
instead signalled by a decrease of the hadron masses proportional to this part
of the condensate evolution. In the nuclear medium the two signatures are simultaneously
present. \bigskip

We acknowledge stimulating discussions with Profs. M. Birse, G. Chanfray, C.
Shakin and H. Toki. P.A.M. Guichon gratefully thanks the CSSM Adelaide for its
support while part of this work was done. This work was supported in part by
the Australian Research Council.

\end{document}